\documentclass[aps,prd,
unsortedaddress,superscriptaddress,showpacs,nofootinbib,10pt,twocolumn]{revtex4-1}
  
\pdfoutput=1 
 
\usepackage{epsfig,amssymb,amsfonts,amsmath,mathtools,bm,color,xcolor,graphicx,braket,adjustbox,esint,upgreek,dcolumn,psfrag,hyperref}
\usepackage{multirow}
\usepackage[utf8]{inputenc}
\usepackage[english]{babel}
\usepackage{slashed}
\usepackage{orcidlink}

\newcommand{\be}{\begin{equation}}
\newcommand{\ee}{\end{equation}}
\newcommand{\bea}{\begin{eqnarray}}
\newcommand{\eea}{\end{eqnarray}}
\newcommand{\beas}{\begin{eqnarray*}}
\newcommand{\eeas}{\end{eqnarray*}}
\newcommand{\ds}{\displaystyle}

\newcommand{\Br}{\mbox{Br}}

\def\vec#1{\boldsymbol{#1}}

\newcommand{\chic}{\chi_{c2}(2P)}

\graphicspath{{Figs.dir/}}

\newcommand{\ijs}{\affiliation{Jozef Stefan Institute, Jamova 39, 1000, Ljubljana, Slovenia}}

\newcommand{\lis}{\affiliation{CeFEMA, Center of Physics and Engineering of Advanced Materials, Instituto Superior T{\'e}cnico, Avenida Rovisco Pais 1, 1049-001 Lisboa, Portugal}}

\newcommand{\rub}{\affiliation{Institut f\"ur Theoretische Physik II, Ruhr-Universit\"at Bochum, D-44780 Bochum, Germany }}

\newcommand{\fzj}{\affiliation{Institute for Advanced Simulation, Institut f\"ur Kernphysik and J\"ulich Center for Hadron Physics, Forschungszentrum J\"ulich, D-52425 J\"ulich, Germany}}

\newcommand{\itp}{\affiliation{CAS Key Laboratory of Theoretical Physics, Institute of Theoretical Physics, \\Chinese Academy of Sciences, Beijing 100190, China}}

\newcommand{\ucas}{\affiliation{School of Physical Sciences, University of Chinese Academy of Sciences, Beijing 100049, China}}

\newcommand{\ific}{\affiliation{Departamento de F\'{\i}sica Te\'orica and IFIC, Centro Mixto Universidad de Valencia-CSIC Institutos de Investigaci\'on de Paterna, Aptdo.22085, 46071 Valencia, Spain}
}

\newcommand{\peng}{\affiliation{Peng Huanwu Collaborative Center for Research and Education, Beihang University, Beijing 100191, China}}

\begin{document}

\title{Production of the $X(4014)$ as the spin-2 partner of $X(3872)$ in $e^+e^-$ collisions}

\author{Pan-Pan Shi\orcidlink{0000-0003-2057-9884}}\email{panpan@ific.uv.es}
\itp \ucas \ific

\author{Vadim Baru\orcidlink{0000-0001-6472-1008}}\email{vadim.baru@tp2.rub.de}
\rub

\author{Feng-Kun~Guo\orcidlink{0000-0002-2919-2064}}\email{fkguo@itp.ac.cn}
\itp \ucas \peng

\author{Christoph~Hanhart\orcidlink{0000-0002-3509-2473}}\email{c.hanhart@fz-juelich.de}
\fzj 

\author{Alexey Nefediev\orcidlink{0000-0002-9988-9430}}\email{Alexey.Nefediev@ijs.si}
\ijs \lis

\begin{abstract}

In 2021, the Belle collaboration reported the first observation of a new structure in the $\psi(2S) \gamma$ final state produced in the two-photon fusion process. In the hadronic molecule picture, this new structure can be associated with the shallow isoscalar $D^*\bar{D}^*$ bound state and as such is an excellent candidate for the spin-2 partner of the $X(3872)$ with the quantum numbers $J^{PC}=2^{++}$ conventionally named $X_2$. In this work we evaluate the electronic width of this new state and argue that its nature is sensitive to its total width, the experimental measurement currently available being unable to distinguish between different options. Our estimates demonstrate that the planned Super $\tau$-Charm Facility offers a promising opportunity to search for and study this new state in the invariant mass distributions for the final states $J/\psi\gamma$ and $\psi(2S)\gamma$.
\end{abstract}

\maketitle

\section{Introduction}

The last two decades have witnessed the discovery of a large number of new hadron structures. Some of them possess the properties, such as the position in the mass spectrum, quantum numbers, and decay width, inconsistent with the predictions of the traditional quark model for mesons and baryons composed as quark-antiquark pairs or three-quark clusters, respectively. These structures are conventionally referred to as exotic states. Many experimental and theoretical studies are aimed to understand their
nature --- for recent reviews with varying foci see, for example, Refs.~\cite{Chen:2016qju,Hosaka:2016pey,Esposito:2016noz,Lebed:2016hpi,Ali:2017jda,Olsen:2017bmm,Guo:2017jvc,Albuquerque:2018jkn,Liu:2019zoy,Guo:2019twa,Brambilla:2019esw,Chen:2022asf}.

The first exotic candidate in the hidden-charm sector, the $X(3872)$ (also known as $\chi_{c1}(3872)$ \cite{ParticleDataGroup:2022pth}), was found by the Belle collaboration in 2003 \cite{Belle:2003nnu}. Since then, a succession of hidden-charm structures have been observed in the collisions with the center-of-mass energy above the open-charm threshold. In 2021, the Belle collaboration reported a hint of the existence of an isoscalar structure with the mass and width
\begin{align}
\begin{split}
M_{X_2}&=(4014.3 \pm 4.0 \pm 1.5)~\mbox{MeV},\\
\Gamma_{X_2}&=(4\pm 11 \pm 6)~\mbox{MeV},
\end{split}\label{X2masswidth}
\end{align}
respectively, observed in the two-photon fusion process~\cite{Belle:2021nuv}. Although the global significance of the new structure is only $2.8\sigma$ (so additional experimental studies would be necessary), the information already collected and conveyed reveals that the mass of this new structure perfectly matches the prediction for the spin-2 partner of the $X(3872)$ based on the heavy quark spin symmetry (HQSS)~\cite{Nieves:2012tt,Guo:2013sya}. In particular, already its residing within just a few MeV from the $D^*\bar{D}^*$ threshold hints towards its molecular interpretation and makes it more promising candidate for the $X(3872)$ spin partner than the well established remote state $\chi_{c2}(3930)$ \cite{ParticleDataGroup:2022pth}. Furthermore, its measured width has the same order of magnitude as the prediction in
Refs.~\cite{Albaladejo:2015dsa,Baru:2016iwj}. Therefore, this narrow state is a potential candidate for a $D^* \bar{D}^*$ molecule with $J^{PC}=2^{++}$ and conventionally tagged as $X_2$ (would be $\chi_{c2}(4014)$ according to the Particle Data Group naming scheme; we, however, choose to stick to the name
proposed originally, $X_2$, to avoid confusion with the generic tensor charmonium $\chic$ also discussed below). Its existence was initially predicted in Ref.~\cite{Tornqvist:1993ng} and then explored in detail in Refs.~\cite{Molina:2009ct,Nieves:2012tt,Hidalgo-Duque:2012rqv,Sun:2012zzd,Tornqvist:1993ng,Hidalgo-Duque:2013pva,Guo:2013sya,Albaladejo:2013aka,Liang:2010ddf,Swanson:2005tn,Albaladejo:2015dsa,Baru:2016iwj,Cincioglu:2016fkm,Baru:2017fgv,Ortega:2017qmg,Mutuk:2018zxs,Wang:2020dgr,Dong:2021juy,Montana:2022inz,Wu:2023rrp,Wang:2023hpp} using various phenomenological approaches.
Alternatively, the new structure reported by the Belle collaboration \cite{Belle:2021nuv} is considered as a $D^* \bar{D}^*$ molecule with $J^{PC}=0^{++}$~\cite{Duan:2022upr,Yue:2022gym}. However, in this case, one would
expect a strong $S$-wave coupling to the $D\bar{D}$ channel for such a state and, therefore, naturally a large total width,
contrary to what is observed at Belle. The compact tetraquark model~\cite{Maiani:2014aja,Wu:2018xdi,Shi:2021jyr,Giron:2021sla} and the conventional $2P$ charmonium picture~\cite{Godfrey:1985xj,Li:2009ad} have been employed to explore the $2^{++}$ states, even though the authors of those works were not primarily focused on the $X_2$ state.

To distinguish between different interpretations of the $X_2$, it is essential to obtain exhaustive experimental information on it, including the quantum numbers and branching fractions for its various decay channels.
So far, the Belle collaboration has only reported the observation of a hint of a new structure near the $D^{*0}\bar{D}^{*0}$ threshold at $M_{\rm th}=(4013.7\pm0.1)$~MeV, however, further properties of this state still remain unclear. To begin with, the large uncertainty in its experimental mass 
does not allow one to firmly conclude whether the state resides below or above the $D^{*0} \bar{D}^{*0}$ threshold. Also, the quantum numbers of the new structure are not yet established, and the uncertainty in its width determination is quite large. Finally, the branching fractions for its different decay channels, such as $D \bar{D}$, $D^* \bar{D}$, $D \bar{D}^* \gamma$ \cite{Albaladejo:2015dsa} and the relation between the decay widths into $J/\psi\gamma$ and $\psi(2S)\gamma$~\cite{Shi:2023mer} remain unknown.

To summarize, while some hint for the $X_2$ candidate has been provided by the Belle collaboration in the two-photon fusion process \cite{Belle:2021nuv}, the substantial uncertainties in its mass and width, as well as the limited statistics, highlight the need for additional theoretical efforts to facilitate further experimental studies of this state. The latter ones, like many other possible searches for hidden-charm exotic states, involve a variety of methods, including their production in prompt $pp$ collisions, decays of $B$ mesons, and the $e^+e^-$ annihilation. On the one hand, exotic states production in the prompt $pp$ collisions is inclusive, and thus its description has to involve simulations with the Monte Carlo event generators \cite{Artoisenet:2009wk,Guo:2014sca,Shi:2021hzm}. On the other hand, the estimation of the branching fraction for the process $B\to D_s \bar{D}^* (D^*_s \bar{D}^*) \to X_2 K$, obtained in Ref.~\cite{Wu:2023rrp} using the effective Lagrangian technique, is about $10^{-5}$. This is an order of magnitude smaller than the experimentally measured branching fraction $(2.1\pm 0.7)\times 10^{-4}$ for the decay $B^+ \to X(3872) K^+$ \cite{ParticleDataGroup:2022pth}. This difference highlights an additional challenge for the searches for the $X_2$ in $B$-meson decays. In this work, we investigate the direct production of the $X_2$ in the $e^+e^-$ annihilation using the vector meson dominance (VMD) model.

This paper is organized as follows. In Sec.~\ref{Sec:Formalism},
we present the interaction Lagrangians and relevant Feynman diagrams contributing to the decay process $X_2\to e^+e^-$.
The predicted $X_2$ electronic width and its possible role in establishing the $X_2$ nature are discussed in Sec.~\ref{Sec:Result}. The direct production of the $X_2$ in $e^+e^-$ collisions at the future high-luminosity Super $\tau$-Charm Facility (STCF) is addressed in Sec.~\ref{Sec:STCF}.
We provide our summary and give an outlook in the concluding Sec.~\ref{Sec:Summary}.

To shorten notations, everywhere throughout the paper, the widths of the two-body decays of the form $\Gamma(a\to bc)$ are denoted as $\Gamma_a^{bc}$ while the total width of the hadron $a$ is denoted as $\Gamma_a$.

\section{Formalism} \label{Sec:Formalism}

\subsection{The Lagrangian and vertices}

\begin{figure*}[t]
\centerline{
\scalebox{0.9}{\epsfig{file=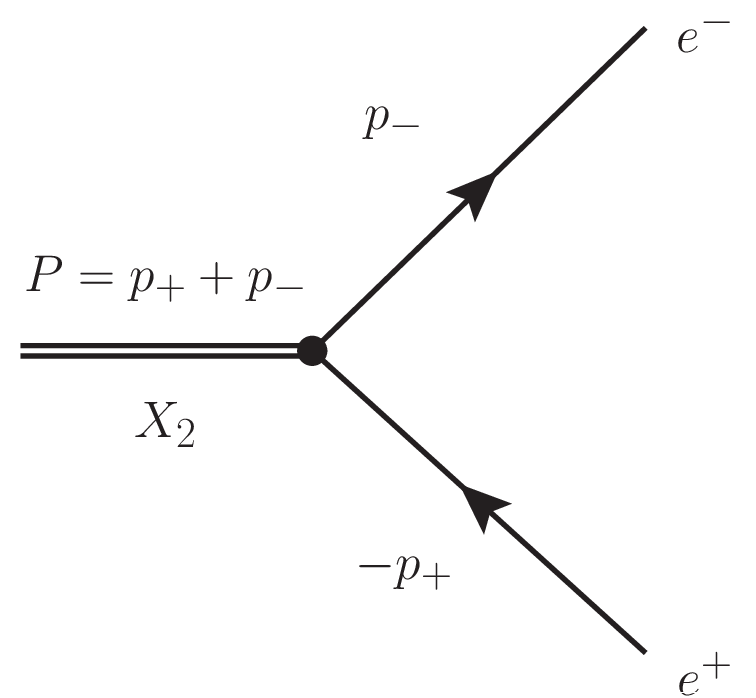,width=0.24\textwidth}
\epsfig{file=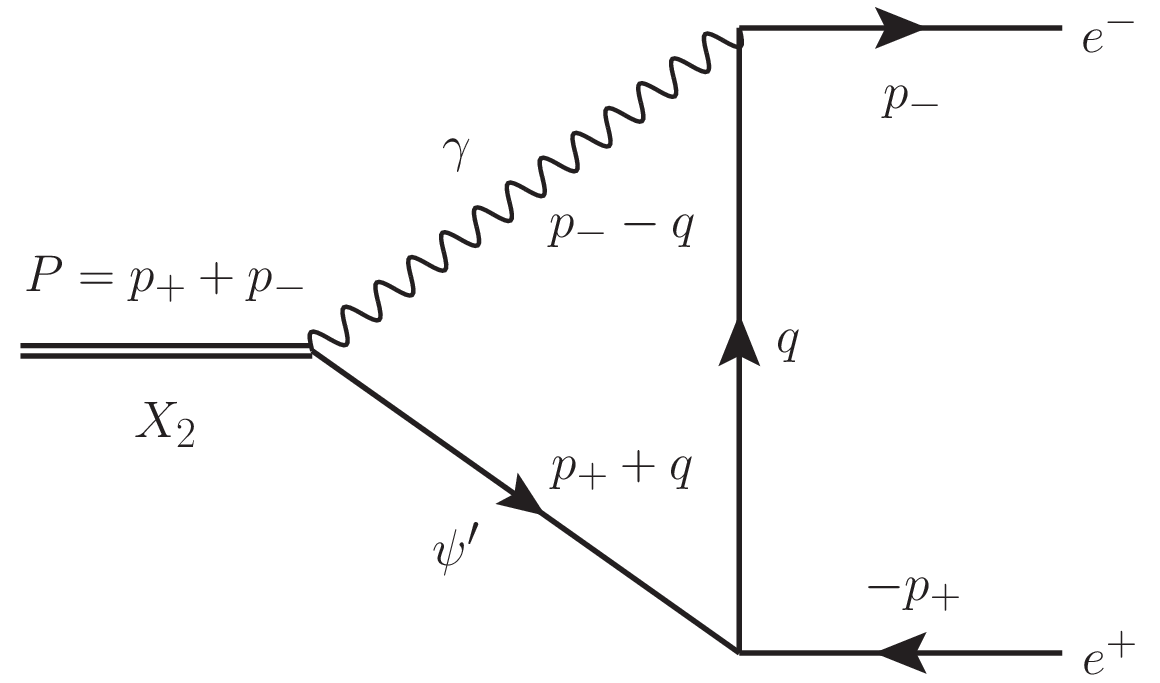,width=0.38\textwidth}
\epsfig{file=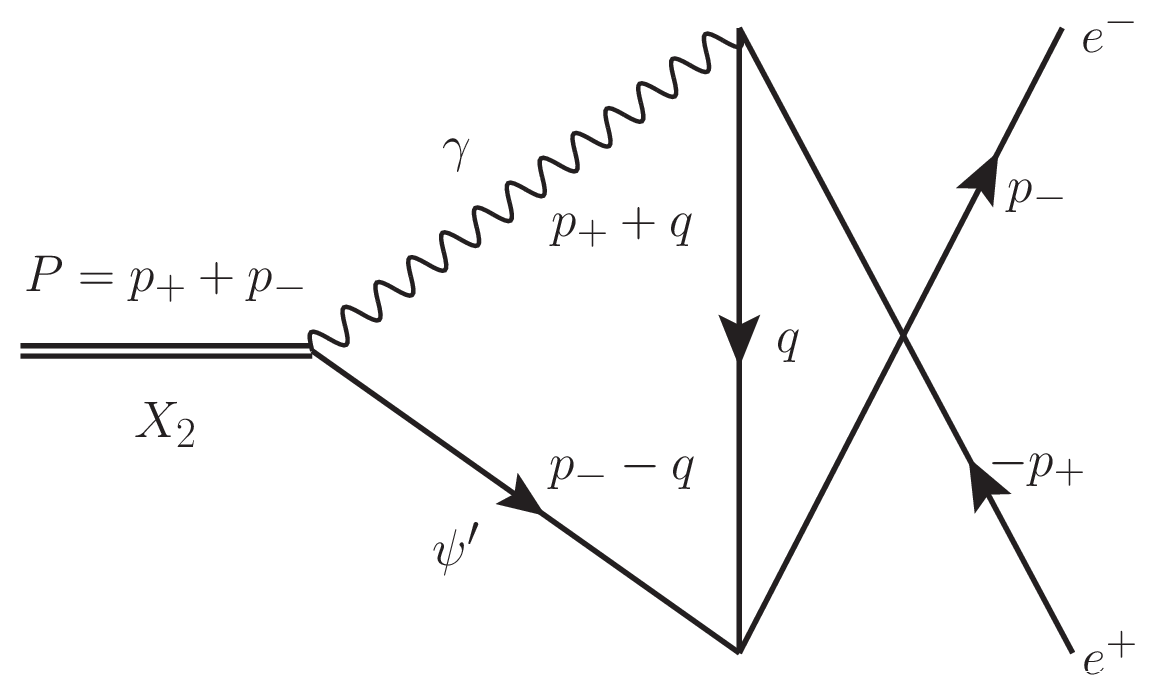,width=0.38\textwidth}}}
\caption{Different contributions to the amplitude of the decay $X_2\to e^+e^-$: the first diagram [diagram (a)] represents short-range contributions and the other two [diagrams (b) and (c)] are for the transition $X_2\to\psi\gamma\to e^+e^-$, with $\psi=J/\psi$, $\psi'$.}\label{Fig:X2epem0}
\end{figure*}

The direct production of the $X_2$ in the $e^+e^-$ collisions through a single photon is forbidden by conservation laws but can occur through a two-photon process $e^+e^-\to \gamma^*\gamma^*\to X_2$. The invariance of the transition amplitude under the time reversal and $P$-parity transformations, called the principle of detailed balance~\cite{Martin:2008zz}, implies that the amplitude for the process $e^+e^- \to X_2$ is identical to that of the decay $X_2\to e^+e^-$, so in what follows we study the latter process. In the VMD approach, the mechanism of the decay $X_2\to e^+e^-$ proceeding via two virtual photons involves the intermediate states $\psi\gamma$ (with $\psi $ for $J/\psi$ and $\psi(2S)$ in the considered energy range), as depicted in Fig.~\ref{Fig:X2epem0}. Assuming the $X_2$ to be a pure $D^*\bar{D}^*$ molecule, the decay of $X_2\to\psi\gamma$ proceeds through the charmed-meson loops \cite{Shi:2023mer}. Here, with this microscopic picture in mind and for the sake of simplicity, we introduce an effective vertex $X_2\to\psi\gamma$. To this end we notice that there are four independent and gauge-invariant structures describing the $\psi\gamma$ final state,
\bea
{\cal S}_{\rho\sigma}^{(1)}&=&
g_{\rho\sigma} (\partial_\alpha F_{\mu\nu}) (\partial^\alpha \psi^{\mu\nu}), \label{Eq:S1}\\
{\cal S}_{\rho\sigma}^{(2)}&=&(\partial_\rho F_{\mu\nu})(\partial_\sigma \psi^{\mu\nu})
+(\partial_\sigma F_{\mu\nu})(\partial_\rho \psi^{\mu\nu})\nonumber\\
&&-\ds\frac12g_{\rho\sigma}(\partial_\alpha F_{\mu\nu})(\partial^\alpha \psi^{\mu\nu}), \label{Eq:S2}\\
{\cal S}_{\rho\sigma}^{(3)}&=&(\partial_\rho\partial_\sigma F_{\mu\nu})\psi^{\mu\nu}
+F_{\mu\nu}(\partial_\rho\partial_\sigma \psi^{\mu\nu}), \label{Eq:S3}\\
{\cal S}_{\rho\sigma}^{(4)}&=&F_{\rho\beta} \psi^{\beta}_\sigma
+ F_{\sigma\beta} \psi^{\beta}_\rho -\ds\frac12g_{\rho\sigma} F_{\mu\nu} \psi^{\mu\nu}\label{Eq:S4},
\eea
where $\psi^{\mu\nu}\equiv\partial^{\mu}\psi^{\nu}-\partial^{\nu}\psi^{\mu}$ and $F^{\mu\nu}$ is the standard electromagnetic field tensor. The above structures are supposed to be contracted with the symmetric $X_2$ polarization tensor $\varepsilon^{\rho\sigma}$ that is subject to the standard constraints,
\be
 P^\rho\varepsilon_{\rho\sigma}(P)=0,\quad P^\sigma\varepsilon_{\rho\sigma}(P)=0,\quad
g^{\rho\sigma}\varepsilon_{\rho\sigma}(P)=0,
\label{Eq:tensprop}
\ee
with $P$ for the $X_2$ 4-momentum. Explicit expression for the polarization tensor of a massive spin-2 particle can be found in Refs.~\cite{Koenigstein:2015asa,Shastry:2021asu}. Then, since the tensor $\varepsilon^{\rho\sigma}$ is traceless, the whole structure ${\cal S}_{\rho\sigma}^{(1)}$ and the last terms in ${\cal S}_{\rho\sigma}^{(2)}$ and ${\cal S}_{\rho\sigma}^{(4)}$ do not contribute to the decay amplitude. Furthermore, since $M_{X_2}-M_\psi\ll M_\psi$, we retain only the structures that survive in the nonrelativistic limit for the vector $\psi$ (the $X_2$ is at rest, so it is always nonrelativistic). This allows us to neglect the remaining parts of the structures ${\cal S}_{\rho\sigma}^{(2)}$ and ${\cal S}_{\rho\sigma}^{(3)}$ since they are proportional to the third power of the final-state 3-momentum. Indeed, due to the condition $P^\rho\varepsilon_{\rho\sigma}(P)=0$
taken in the $X_2$ rest frame one concludes that only the spatial components $\varepsilon_{ij}$ of the $X_2$ polarization tensor survive and, therefore, so do only the spatial components of the 4-derivatives in Eqs.~\eqref{Eq:S2} and \eqref{Eq:S3}. Such contributions are suppressed, and we disregard them. Therefore, only the first two terms in the structure ${\cal S}_{\rho\sigma}^{(4)}$ should be retained, and the corresponding effective interaction Lagrangian reads
\be
{\cal L}_{X_2\psi\gamma}=g_{X_2\psi\gamma}X_{2}^{\rho\sigma}F_{\sigma\beta}\psi_\rho^\beta,
\label{Lag}
\ee
where the relativistic coupling $g_{X_2\psi\gamma}$ has the dimension $m^{-1}$. The vertex $X_{2\sigma\rho}(P)\to \gamma_{\alpha}(k_1)\psi_{\beta}(k_2)$ derived from the Lagrangian \eqref{Lag} is
\begin{align}
\Gamma_{\sigma\rho\alpha\beta}^{\psi\gamma}(k_1,k_2)=&-ig_{X_2\psi\gamma}\Bigl[k_{1\sigma}k_{2\alpha}g_{\beta\rho}-k_{1\sigma}k_{2\rho}g_{\alpha\beta}\nonumber\\
&+k_{1\beta}k_{2\rho}g_{\sigma\alpha}-(k_1\cdot k_2) g_{\alpha\sigma}g_{\beta\rho}\Bigr].
\label{Eq:vertex_gamma_psi}
\end{align}

The subsequent transition $\psi\to\gamma^*\to e^+e^-$ is described by the VMD Lagrangian \cite{OConnell:1995nse,Klingl:1996by,Korchin:2011ze}
\be
{\cal L}_{\psi\gamma}=-\frac{e}{2}\frac{f_{\psi}Q_{c}}{M_{\psi}}F^{\mu\nu}\psi_{\mu\nu},
\label{Eq:VMD}
\ee
where $Q_{c}=3/2$ is the electric charge of the charm quark and $f_\psi$ represents the corresponding decay constant of $\psi$.\footnote{This so-called second representation of VMD~\cite{Bauer:1977iq} employed here arises from the resonance chiral theory~\cite{Ecker:1988te} and is $U(1)$ gauge invariant unlike the first representation of VMD~\cite{Sakurai:1969,Meissner:1987ge,OConnell:1995nse}.}

The process $\gamma^*\to e^+e^-$ is described by the standard Lagrangian of quantum electrodynamics, which yields the vertex $\psi_{\mu}(k)\to e^+(p_1) e^-(p_2)$ in the form
\begin{align}
\Gamma^{e^+e^-}_\mu(k) = -ie^2\frac{f_{\psi}Q_c}{M_{\psi}}\left(g_{\mu\nu}-\frac{k_{\mu}k_{\nu}}{M_{\psi}^2}\right) \gamma^{\nu}.
\label{Eq:VMD_vertex}
\end{align}

\subsection{The electronic $X_2$ width}

With the explicit form of the vertices derived in the above section, we are in a position to derive the contribution to the decay amplitude $X_2\to e^+e^-$ from the triangle loop diagrams (b) and (c) depicted in Fig.~\ref{Fig:X2epem0},
\begin{align}
i {\cal M}_{X_2 \to e^+e^-}^\text{loop}&\equiv \epsilon^{*\sigma\rho}{\cal M}_{\sigma \rho}^\text{loop}\nonumber\\[-2mm]
\label{Eq:loop_function}\\[-2mm]
&=\epsilon^{*\sigma\rho}\sum_{\psi=J/\psi,\psi(2S)}\left(J^{(b)\psi}_{\sigma\rho}+J^{(c)\psi}_{\sigma\rho}\right),\nonumber
\end{align}
where
\begin{align}
J^{(b)\psi}_{\sigma\rho}=&\int\frac{d^4q}{(2\pi)^4}S_{\psi}^{\mu\nu}(p_++q)S_{\gamma}^{\alpha\beta}(p_--q)\nonumber\\
&\times\Gamma_{\sigma\rho\beta\nu}^{\psi\gamma}(p_--q,p_++q)\nonumber\\
&\times\bar u(p_-)\gamma_{\alpha}S_F(q)\Gamma_{\mu}^{e^+e^-}(p_++q)v(p_+),\nonumber\\[-2mm]
\label{Js}\\[-2mm]
J^{(c)\psi}_{\sigma\rho}=&\int\frac{d^4q}{(2\pi)^4}S_{\psi}^{\mu\nu}(p_--q)S_{\gamma}^{\alpha\beta}(p_++q)\nonumber\\&\times\Gamma_{\sigma\rho\beta\nu}^{\psi\gamma}(p_+ +q,p_- - q)\nonumber\\
&\bar u(p_-)\Gamma_{\mu}^{e^+e^-}(p_--q)S_F(q)\gamma_{\alpha}v(p_+),\nonumber
\end{align}
and
\begin{align}
&S_{\gamma}^{\alpha\beta}(k)=\frac{-ig^{\alpha\beta}}{k^2+i\epsilon},\nonumber\\
&S_{\psi}^{\mu\nu}(k)=\frac{i}{k^2-m_{\psi}+i\epsilon}\left(-g_{\mu\nu}+\frac{k_{\mu}k_{\nu}}{m_{\psi}^2}\right),
\\
&S_F(q)=\frac{i\slashed{q}}{q^2+i\epsilon}\nonumber
\end{align}
for the photon, charmonium, and fermion propagator, respectively; the electron mass is neglected.

Since the loop integrals in the amplitude \eqref{Eq:loop_function} are ultraviolet (UV) divergent, we introduce a counterterm (the diagram (a)
in Fig.~\ref{Fig:X2epem0}) that, in the covariant orbital-spin scheme~\cite{Zou:2002yy}, takes the form
\begin{align}
i{\mathcal M}^{\text{cont}}\equiv&\,
\epsilon^{*\sigma\rho}{\cal M}_{\sigma \rho}^\text{cont}\nonumber\\
=&\,
i\lambda_P \epsilon^{*\sigma\rho} \left[ r_{\sigma} \bar{u}(p_-)\gamma_{\rho}v(p_+) + r_{\rho} \bar{u}(p_-)\gamma_{\sigma}v(p_+)\right] \nonumber\\
&+i\lambda_F \epsilon^{*\sigma\rho} \Bigl[r_{\sigma}r_{\rho}r_{\alpha} \bar{u}(p_-)\gamma^{\alpha}v(p_+)
\label{Eq:amp_contact}\\
&+\frac15 M_{X_2}\left(g_{\rho\alpha}r_{\sigma}+g_{\sigma\alpha}r_{\rho}\right)\bar{u}(p_-)\gamma^{\alpha}v(p_+) \Bigr],\nonumber
\end{align}
where $r^{\mu}=p_+^{\mu}-p_-^{\mu}$, and $\lambda_P$ and $\lambda_F$ are the low-energy constants (LECs) parameterizing the $P$- and $F$-wave contributions, respectively.

In addition to the contributions from the intermediate states $\psi\gamma$ ($\psi=J/\psi$, $\psi(2S)$), the decay $X_2\to e^+e^-$ can also proceed as $X_2\to \psi V\to e^+e^-$, with $V=\rho$, $\omega$. Meanwhile, in the decay $X(3872)\to e^+e^-$ the contributions from $\psi V$ in the intermediate state were found to be significantly smaller than those from $\psi\gamma$ \cite{Denig:2014fha}. The same hierarchy of the contributions from $\psi\gamma$ and $\psi V$ is also expected for the $X_2$ regarded as the $X(3872)$ spin partner. We, therefore, neglect the contributions from the $\psi\rho$ and $\psi\omega$ intermediate states to the total amplitude of the decay $X_2\to e^+e^-$.

Then we finally arrive at the electronic decay width of the $X_2$ in the form
\begin{align}
\Gamma_{X_2}^{ee}=\frac{|{\vec p}_-|}{40\pi M_{X_2}^2}{\cal M}_{\sigma\rho}P^{\sigma\rho\sigma'\rho'}(P,M_{X_2}){\cal M}_{\sigma'\rho'}^*,
\label{Eq:X2_width}
\end{align}
where, in neglect of the electron mass, the momenta of the electron and positron in the center-of-mass frame are $|{\vec p}_-|=|{\vec p}_+|=M_{X_2}/2$, ${\cal M}_{\sigma\rho} = {\cal M}_{\sigma\rho}^\text{loop} + {\cal M}_{\sigma\rho}^\text{cont}$, and $P^{\sigma\rho\sigma'\rho'}(P,M_{X_2})$ is the $X_2$ density tensor,
\begin{align}
P^{\sigma\rho\sigma'\rho'}(P,M_{X_2})=&\,\frac12\bar{g}_{\sigma\sigma'}(P,M_{X_2})\bar{g}_{\rho\rho'}(P,M_{X_2})\nonumber\\
&+\frac12\bar{g}_{\sigma\rho'}(P,M_{X_2})\bar{g}_{\rho\sigma'}(P,M_{X_2})\\\
&-\frac{1}{3}\bar{g}_{\sigma\rho}(P,M_{X_2})\bar{g}_{\sigma'\rho'}(P,M_{X_2}),\nonumber
\end{align}
with
\be
\bar g_{\mu\nu}(P,M_{X_2}) \equiv -g_{\mu\nu}+\frac{P_{\mu}P_{\alpha}}{M_{X_2}^2}.
\ee

\subsection{ Comment on the two-photon decay of $X_2$}

Under the hypothesis of the quantum numbers $2^{++}$ of the resonance observed by Belle in the two-photon fusion process, its partial decay width was measured to be \cite{Belle:2021nuv}
\be
\Gamma_{X_2}^{\gamma\gamma}\Br(X_2\to\psi(2S)\gamma)=(1.2\pm 0.4\pm 0.2)~\mbox{eV},
\label{Eq:widthexp}
\ee
where $\Gamma_{X_2}^{\gamma\gamma}$ is the two-photon decay width of the $X_2$. Therefore, in order to extract $\Br(X_2\to\psi(2S)\gamma)$ separately, one needs a theoretical estimate for $\Gamma_{X_2}^{\gamma\gamma}$. 
In Ref.~\cite{Baru:2017fgv}, an order-of-magnitude estimate for the decay of the $X(3915)$ was made under the assumption for this state to be a tensor $D^*\bar{D}^*$ molecule---a spin-2 partner of the $X(3872)$
\be\label{X2gg}
\Gamma(X_2\to \gamma\gamma) \simeq 0.1 ~\mbox{keV}.
\ee
We note that the actual assignment for the state previously known as the $X(3915)$ is still obscure: although the Particle Data Group tends to assign it to a scalar state $\chi_{c0}$~\cite{ParticleDataGroup:2022pth}, its spin-2 interpretation is advocated, for example, in Ref.~\cite{Ji:2022vdj}. Thus, as a benchmark, we mention here a generic $\bar{c}c$ spin-2 charmonium $\chi_{c2}$ lying in the studied energy range and stick to the quark model estimate for it provided in Ref.~\cite{A:2023bxv},
\be\label{chic2gg}
\Gamma(\chi_{c2}\to \gamma\gamma) \simeq 1 ~\mbox{keV}
\ee
which exceeds the result (\ref{X2gg}) obtained for the molecule by an order of magnitude.

Since the result (\ref{X2gg}) depends on the mass of the decaying state, we repeated the calculations of Ref.~\cite{Baru:2017fgv} for the mass $M_{X_2}=4014$~MeV to find for the finite and scheme-independent helicity-0 contribution to the two-photon decay width of the $X_2(4014)$,
\be
\Gamma_{X_2}^{\gamma\gamma}[\mbox{hel-0}]= %0.15*(g[GeV])^2 eV
0.15\Bigl(g_{X_2D^*\bar{D}^*}[\mbox{GeV}]\Bigr)^2~\mbox{eV},
\label{X2gg0}
\ee
where the $X_2$ coupling to the constituents can be calculated as \cite{Landau,Weinberg:1965zz}
\be
\frac{g_{X_2D^*\bar{D}^*}^2}{4\pi}=32m_{D^*}\sqrt{m_{D^*}E_B},
\label{Eq:gx2}
\ee
with
\be
E_B=2m_{D^*}-M_{X_2},
\label{EBdef}
\ee
and $m_{D^*}$ for the $D^*$ mass. 
Although the existing measurement \eqref{X2masswidth} is rather uncertain and does not allow one to precisely fix the binding energy of the $X_2$, using $E_B$ 
around 5 MeV as an upper bound, one can find
\be
\Gamma_{X_2}^{\gamma\gamma}[\mbox{hel-0}]\lesssim 12~\mbox{eV},
\label{hel0}
\ee 
that complies very well with the results of Ref.~\cite{Baru:2017fgv}. It is therefore expected that also for the $X_2(4014)$ the helicity-2 component would prevail in the two photon decays and the numerical result (\ref{X2gg}) is valid as an order-of-magnitude estimate also for the $X_2(4014)$.
Then, in the absence of any 
information on the $X_2(4014)$ decays,  
in what follows, we resort to Eq.\,\eqref{X2gg} for the two-photon decay width of the $X_2(4014)$ under the assumption of its molecular nature.

\subsection{Parameters estimation}

The electronic decay width of the $X_2$ from Eq.\,\eqref{Eq:X2_width} 
depends on several unknown parameters, namely (i) the two LECs $\lambda_P$ and $\lambda_F$, (ii)
the effective coupling constant $g_{X_2\psi\gamma}$ [see Eq.\,\eqref{Eq:vertex_gamma_psi}], and (iii) the decay constants $f_\psi$ [see Eq.\,\eqref{Eq:VMD_vertex}] for $\psi=J/\psi$ and $\psi(2S)$. The LECs will be discussed in Sec~\ref{Sec:Result} while here we focus on the extraction of the couplings from the data.

The results reported in Ref.~\cite{Shi:2023mer} imply that, as an order-of-magnitude estimate, one has
\be
\Gamma_{X_2}^{J/\psi\gamma}\simeq
\Gamma_{X_2}^{\psi(2S)\gamma}.
\label{jpsi}
\ee
The latter width (hereinafter tagged as $\Gamma_{X_2}^{\psi\gamma}$) can be extracted from the measurement \eqref{Eq:widthexp} as
\be
\Br(X_2\to\psi\gamma)=\frac{\Gamma_{X_2}^{\psi\gamma}}{\Gamma_{X_2}}\simeq 10^{-2},
\label{Eq:X2psigam}
\ee
where we used the estimate \eqref{X2gg} and took the right-hand side of Eq.\,\eqref{Eq:widthexp} as $\simeq1$~eV.

On the other hand, the vertex defined in Eq.\,\eqref{Eq:vertex_gamma_psi} gives
\begin{align}
\Gamma_{X_2}^{\psi\gamma}=&\,\frac{g_{X_2\psi\gamma}^2}{960\pi M_{X_2}^7}\left(6M_{X_2}^4+3M_{\psi}^2M_{X_2}^2+M_{\psi}^4\right)\nonumber\\
&\times\left(M_{X_2}^2-M_{\psi}^2\right)^3
\label{Eq:X2_gamma_psi}
\end{align}
and, therefore, we can find the values of the couplings,
\begin{align}
\begin{split}
   &g_{X_2J/\psi\gamma}\simeq 3.6~\mbox{GeV}^{-3/2}\sqrt{\Gamma_{X_2}},\\
&g_{X_2\psi(2S)\gamma}\simeq 0.9~\mbox{GeV}^{-3/2}\sqrt{\Gamma_{X_2}}.
\end{split} \label{g1}
\end{align}

Finally, the decay constants $f_{J/\psi}$ and $f_{\psi(2S)}$ can be extracted directly from the electronic widths of the $J/\psi$ and $\psi(2S)$. To this end we employ the effective vertex from Eq.\,\eqref{Eq:VMD_vertex} to write the amplitude of the decay $\psi_{\mu}(p) \to e^-(k_1)e^+(k_2)$,
\be
{\cal M}_{\psi}=-ie^2\frac{f_{\psi}Q_c}{M_{\psi}}\left(g_{\mu\nu}-\frac{p_{\mu}p_{\nu}}{M_{\psi}^2}\right)\bar u(k_1)\gamma^{\nu} v(k_2)\epsilon^{\mu}(p),
\ee
and arrive at the electronic width
\be
\Gamma_\psi^{ee}=\frac{4\pi\alpha^2}{3M_\psi}f_\psi^2Q_c^2.
\label{Eq:width_psi}
\ee

Then for the measured electronic widths \cite{ParticleDataGroup:2022pth},
\be
\Gamma_{\psi(2S)}^{ee}=2.44~\mbox{keV},\quad\Gamma_{J/\psi}^{ee}=5.53~\mbox{keV},
\ee
one readily finds
\be
f_{J/\psi}=415.49~\mbox{MeV},\quad
f_{\psi(2S)}=294.35~\mbox{MeV}.
\ee

\section{Numerical Results} 
\label{Sec:Result}

We use the dimensional regularization scheme to treat the UV divergent integrals in Eq.\,\eqref{Js}.
In particular, we adopt the $\overline{\text{MS}}$ subtraction scheme and employ the Wolfram Mathematica packages FeynCalc~\cite{Shtabovenko:2020gxv} and FeynHelpers~\cite{Shtabovenko:2016whf}.

After renormalization of the theory, the dependence of the loop amplitudes on the renormalization
scale $\mu$ should be
compensated by the $\mu$-dependence of the short-range constants $\lambda_P$ and $\lambda_F$ from Eq.\,\eqref{Eq:amp_contact}. Lacking an experimental observable that could be used to fix the finite part of the renormalized amplitude, we resort to the method from Ref.~\cite{Guo:2014taa}. Namely, we set $\lambda_P=\lambda_F=0$ and estimate the size of the short-range contribution by varying the scale $\mu$ in a large range from 2.0 to 6.0 GeV, that is, by $M_{X_2}/2\approx 2$~GeV around the central value of $\mu=M_{X_2}\approx 4$~GeV. The obtained values of the loop contribution [diagrams (b) and (c) in Fig.~\ref{Fig:X2epem0}] to the branching fraction of the $X_2\to e^+e^-$ decay are listed in Table~\ref{Tab:X_2_mu}.
From these results one can conclude that the $\mu$-dependence of $\Br_{\rm loop}[X_2\to e^+e^-]$ is relatively mild and so must be the $\mu$-dependence of the contact term. 
Conjecturing that the contact term may be of the same order as the variation observed in Table~\ref{Tab:X_2_mu}, we stick to few units times $10^{-9}$ as on order-of-magnitude estimate for $\Br[X_2\to e^+e^-]$.
For convenience, in the first row of Table~\ref{Tab:X_2Br} we summarize the order-of-magnitude estimates for the branching fractions of various decays of the $X_2$, treated as spin-2 molecule, that were obtained and used in this work.

It is instructive to compare the result obtained above for the molecular $X_2$ with the estimates found in the literature for the generic charmonium {with the quantum numbers $J^{PC}=2^{++}$}. In particular,
\begin{align}
&\Gamma_{\chi_{c2}(1P)}^{ee}=0.014~\mbox{eV (VMD) \cite{Kuhn:1979bb}},\nonumber\\[-2mm]
\label{chi2}\\[-2mm]
&\Gamma_{\chi_{c2}(1P)}^{ee}=0.07~\mbox{eV (NRQCD) \cite{Kivel:2015iea}},\nonumber
\end{align}
where in parentheses we quote the method used in the calculations, with NRQCD for nonrelativistic quantum chromodynamics. {The quark model predicts the wave function at the origin to take similar numerical values for both $\chi_{c2}(1P)$ and its first radial excitation $\chic$~\cite{Eichten:1995ch}, so we expect  
$\Gamma_{\chic}^{ee}$ to be close to the values quoted in Eq.\,\eqref{chi2}}. Since the total width $\Gamma_{\chic}$ is expected to be of the order of several dozen MeV \cite{ParticleDataGroup:2022pth}, from Eq.\,\eqref{chi2} we conclude that $\Br[\chic\to e^+e^-]$ also constitutes a few units times $10^{-9}$, like for the $X_2$ as a $2^{++}$ molecule. Meanwhile, we notice an important difference between these two results: while the above estimate for the $\chic$ relies on the natural expectation that $\Gamma_{\chic}\sim 10$~MeV,
the result for the $X_2$ as a molecule is valid for \emph{any} $\Gamma_{X_2}$ compatible with the measurements \eqref{X2masswidth} and \eqref{Eq:widthexp} --- see the scaling $g_{X_2\psi\gamma}\propto\sqrt{\Gamma_{X_2}}$ in Eq.\,\eqref{g1}. 
This might bring important insights on the nature of the $X_2$ state.
Indeed, according to the measurement \eqref{X2masswidth}, the width $\Gamma_{X_2}$ cannot be much bigger than 10~MeV. 
If the total width $\Gamma_{X_2}$ is small (of the order 1~MeV or less), then the $X_2$ is only consistent with a predominantly $D^*\bar{D}^*$ molecular structure.
Notice that the rate of the decay of a molecular state is proportional to $g_{X_2D^*\bar{D}^*}^2$, and thus proportional to the square root of the binding energy; see Eq.\eqref{Eq:gx2}. 
Therefore, the electronic width of the $X_2$ in the molecular picture probes both the mass of the $X_2$ and its width. 
Furthermore, in this case, not only the total width looks abnormally small for a generic charmonium, but also the di-electron width of the molecular $X_2$ would be $\Gamma_{X_2}\cdot \Br[X_2\to e^+e^-]\sim 1~\mbox{MeV}\cdot 10^{-9}\sim 10^{-3}~\mbox{eV}$, which is one to two orders of magnitude smaller than the estimates from Eq.\,\eqref{chi2} for the generic $2^{++}$ charmonium. 
On the contrary, as the width $\Gamma_{X_2}\sim 10$ MeV would be consistent with both scenarios, no definite conclusions are possible. On the one hand,
the wave function of the $X_2$ is likely to possess a sizable (that we are unable to quantify at the present level of accuracy) short-range component.
In particular, it could be related with ordinary charmonia, since
quark models predict the generic charmonium state $\chic$ at approximately 3.93...3.95~GeV~\cite{Ebert:2002pp,Ebert:2013iya,Ferretti:2013faa}, that is, within the same energy range from the $X_2(4014)$ as the $\chi_{c1}'$ generic charmonium resides from the $X(3872)$. 
On the other hand, a sizable width of a near-threshold state extracted using the Breit-Wigner parameterization is also not in contradiction with a molecule. For example, the $Z_b(10610)$ and $Z_b(10650)$ discovered by the Belle collaboration~\cite{Adachi:2011mks} have the Breit-Wigner widths of about 18 MeV and 12 MeV, respectively. Nevertheless, their pole positions extracted in a way consistent with unitarity and analyticity are located near the open-bottom thresholds and are fully in line with the molecular nature of the $Z_b$'s \cite{Wang:2018jlv,Baru:2019xnh}.

To assess the uncertainties of the result just obtained we check their dependence on the values of the most essential parameters they depend on. The experimental uncertainty in the $X_2$ mass determination mainly affects the results through the binding energy (see, in particular, Eqs.~\eqref{X2gg0}-\eqref{EBdef}), for which the experiment fails to establish a sizeable lower bound. Then, treating the binding energy in a broad sense as the distance from the resonance pole to the threshold, we estimate the lower bound of the binding energy by the width $\Gamma_{X_2}$. Since, as a conservative estimate for the molecule, we take $\Gamma_{X_2}\simeq 1$~MeV (\emph{c.f.} $E_B=5$~MeV used to reach the estimate in Eq.\,\eqref{hel0}), the absolute value of the $X_2$ dielectron width may decrease by a factor 2 or so. However, by virtue of the relations from Eq.\,\eqref{g1}, the branching fractions $\Br[X_2\to e^+e^-]$ quoted in Tables~\ref{Tab:X_2_mu} and \ref{Tab:X_2Br} will remain intact. The largest uncertainty of the obtained results is expected to come from the variation of the renormalization scale $\mu$ as given in Table~\ref{Tab:X_2_mu}. In addition, according to Ref.~\cite{Shi:2023mer}, the ratio $\Gamma_{X_2}^{\psi(2S)\gamma}/\Gamma_{X_2}^{J/\psi\gamma}$ can deviate from unity within approximately 20\%, which
would propagate to the same uncertainty in  the coupling $g^2_{X_2\psi(2s)\gamma}$. This variation of the coupling constant has  however a much smaller impact on the results than the $\mu$ dependence discussed above. In what follows, to estimate the production cross section of the $X_2$ in $e^+e^-$ collisions, we employ the lower value of $\Br[X_2\to e^+e^-]$ contained in Table~\ref{Tab:X_2_mu}.

Finally, for the sake of comparison, in the last row of Table~\ref{Tab:X_2Br}, we summarize the order-of-magnitude estimates for the branching fractions of a generic $\chi_{2c}(2P)$ charmonium following from Eqs.~\eqref{Eq:widthexp}, \eqref{chic2gg}, \eqref{chi2} and the assumption that relation \eqref{jpsi} approximately holds for generic tensor charmonia, too.

\begin{table}[t!]
\caption{The contribution of the loop amplitudes (diagrams (b) and (c) in Fig.~\ref{Fig:X2epem0}) to the branching fraction of the $X_2\to e^+e^-$ decay evaluated in the $\overline{\text{MS}}$ subtraction scheme for different values of the scale $\mu$.}
\begin{ruledtabular}
\begin{tabular}{lccc}
$\mu$ [GeV] & 2.0 & 4.0 & 6.0 \\
\hline
$\Br_{\rm loop}[X_2\to e^+e^-]\times 10^9$ & 2 & 7 & 11
\end{tabular}
\end{ruledtabular}
\label{Tab:X_2_mu}
\end{table}

\begin{table}[t!]
\caption{ The order-of-magnitude estimates for the branching fractions of various decays of the $X_2$, as a $D^*\bar{D}^*$ molecule (first row) and generic $\bar{c}c$ quarkonium (second row), obtained and employed in this work. The two values quoted for the two-photon decay of the molecule correspond to $\Gamma_{X_2}=1$~MeV and $\Gamma_{X_2}=10$~MeV, respectively.}
\begin{ruledtabular}
\begin{tabular}{lcccc}
Channel & $J/\psi\gamma$ & $\psi(2S)\gamma$ & $\gamma\gamma$ & $e^+e^-$\\
\hline
$\vphantom{\Bigl(}(D^*\bar{D}^*)_{J=2}$ & $10^{-2}$ & $10^{-2}$ & $10^{-4}/10^{-5}$ & $10^{-9}$\\
$\chi_{c2}(2P)$ & $10^{-3}$ & $10^{-3}$ & $10^{-4}$ & $10^{-9}$
\end{tabular}
\end{ruledtabular}
\label{Tab:X_2Br}
\end{table}

\section{Production of $X_2$ at STCF}
\label{Sec:STCF}

In this section we provide estimates for the direct production of the $X_2$ in the $e^+e^-$ collisions. The obtained results should facilitate further searches and studies of the $X_2$ in the experiments at the planned STCF~\cite{Achasov:2023gey}.

Employing the principle of detailed balance discussed above, the difference between the production and decay processes for the $X_2$ lies in the phase space. Consequently, the production of the $X_2$ can be expressed as the partial width for $X_2\to \gamma^*\gamma^*\to e^+ e^-$ multiplied by the appropriate phase-space factor. Then the cross section of the $X_2$ direct production in $e^+e^-$ annihilation can be estimated as (see Appendix A of Ref.~\cite{Shi:2021hzm} for further details)
\begin{align}
\sigma_{C}
\simeq\frac{{\cal M}_{\sigma\rho}P^{(2)}{}^{\sigma\rho\sigma'\rho'}(P,M_{X_2}){\cal M}_{\sigma'\rho'}^*}{4\Gamma_{X_2} \sqrt{s \lambda( s,m_{e^+}^2,m_{e^-}^2)}},
\label{Eq:section_2_1}
\end{align}
where the factor 1/4 stems from averaging over the spins of the initial fermions, $s=(E_{e^+}+E_{e^-})^2=M_{X_2}^2$, and $\lambda(x,y,z)\equiv x^2+y^2+z^2-2xy-2yz-2zx$ is the standard K{\"a}ll\'en triangle function. This formula is derived in the narrow-width approximation consistent with the measured $X_2$ width quoted in Eq.\,\eqref{X2masswidth} above.
Then, with the help of Eq.\,\eqref{Eq:X2_width}, one arrives at the estimate
\be
\sigma_C\simeq \frac{20\pi\Gamma_{X_2}^{ee}}{\Gamma_{X_2} M_{X_2}^2}=\frac{20\pi}{ M_{X_2}^2}\Br [X_2\to e^+e^-] {\simeq 7}~\mbox{pb},
\ee
where we used the value of $\Br [X_2\to e^+e^-]$ from Table~\ref{Tab:X_2Br}.\footnote{For the branching fractions from Table~\ref{Tab:X_2Br}, $\sigma_C\simeq 3,~7,~10$~pb, respectively, with the mean value around 7~pb.}

During the period from 2011 to 2014, the BESIII experiment accumulated an integrated luminosity of around 53 pb$^{-1}$ at the center-of-mass energy $\sqrt{s}=4.090$~GeV~\cite{BESIII:2015qfd,BESIII:2022dxl}. Then, using the known values of the branching fractions~\cite{ParticleDataGroup:2022pth}
\begin{align}
\begin{split}
&\Br[\psi(2S)\to \pi^+\pi^- J/\psi]\simeq(34.68 \pm 0.30) \%, \\
&\Br [J/\psi\to \ell^+\ell^-]\simeq(11.93\pm 0.07)
\%\quad (\ell=e,~\mu),
\end{split} \label{eq:brpsi}
\end{align}
one anticipates that BESIII, working at the center-of-mass energy around 4.014 GeV, can {hardly} detect directly produced $X_2$ events in the $J/\psi\gamma$ and $\psi(2S)\gamma$ invariant mass distributions with {the $J/\psi$'s reconstructed from lepton pairs}. 
% It is, therefore, challenging to search for the $X_2$ at BESIII using the direct production process. 
In the meantime, given the expected integrated luminosity of the STCF around 1~ab$^{-1}$/year~\cite{Achasov:2023gey}, one can expect a considerable amount of approximately {$ 7\,{\rm pb}\times 1\,{\rm ab}^{-1} = 7\times 10^6$} directly produced $X_2$ and, depending on its nature, from $\mathcal{O}(10^2)$ to $\mathcal{O}(10^3)$ reconstructed events in the $J/\psi\gamma$ or $\psi(2S)\gamma$ invariant mass distributions annually, estimated using the values of the branching fraction $\Br[X_2\to \psi\gamma]$ from Table~\ref{Tab:X_2Br}) and those in Eq.\,\eqref{eq:brpsi}.\footnote{The dominant decay modes of the $X_2$ should be the open-charm meson pairs $D\bar D, D\bar D^*+c.c.$ in $D$-wave~\cite{Albaladejo:2015dsa,Baru:2016iwj}. However, the detection efficiency of each $D$ or $\bar D$ meson is at the level of a few to ten per cent. Thus, such modes are not compellingly better options for the search of the $X_2$.} Therefore, possible future studies of the $X_2$ through its direct production at the STCF look promising.\footnote{Of course the $X_2$ can also be produced at both BESIII after the planned BEPCII upgrade~\cite{BESIII:2020nme} and STCF in association with a light vector meson, that is, $e^+e^-\to X_2 \rho^0/\omega$, at the center-of-mass energies $\gtrsim 5$~GeV.}

~

\smallskip

\section{Summary}
\label{Sec:Summary}

In 2021, the Belle collaboration reported an evidence of the presence of an isoscalar structure in the $\psi(2S)\gamma$ invariant mass distribution produced in the two-photon fusion process near the $D^*\bar{D}^*$ threshold \cite{Belle:2021nuv}. This structure is an excellent candidate for the $D^*\bar{D}^*$ molecule conventionally referred to as $X_2$ and predicted using HQSS as the tensor spin partner of the $X(3872)$ \cite{Nieves:2012tt,Guo:2013sya}. However, due to the present limited statistics and significant uncertainty in the $X_2$ pole position, further experimental efforts are required to confirm the existence of this state and reliably investigate its properties.

In this work, we study the direct production of the $X_2$ in $e^+e^-$ collisions and demonstrate that a self-consistent description of the $X_2$ as a tensor $D^*\bar{D}^*$ molecule is achieved if its total width is of the order $\Gamma_{X_2}\simeq 1$~MeV. On the contrary, a clear-cut conclusion about the structure of the $\Gamma_{X_2}$ is not possible if
the width $\Gamma_{X_2}\simeq 10$~MeV. The full wave function of a state in this case can include both a molecular and substantial short-range non-molecular component.
Since both scenarios with a small and a sizable width are consistent with the only currently available Belle measurement \eqref{X2masswidth}, we argue that further experimental investigations of this state are important and timely. To facilitate the corresponding experimental efforts we provide relevant order-of-magnitude estimates for the expected number of the $X_2$ events to be collected in electron-positron collisions in the present experiment BESIII and at the future STCF. We conclude that,
while searches for the $X_2$ candidates at BESIII is challenging, the STCF will offer a new and unique opportunity for its discovery and studies. Finally, we propose that the $J/\psi \gamma$ final state is used in the future studies for the $X_2$.
\medskip

\begin{acknowledgments}

This work is supported in part by the National Natural Science Foundation of China Grant Nos.~12070131001, 12125507, 11835015, and
12047503, the Deutsche Forschungsgemeinschaft (DFG) through the funds provided to the Sino-German Collaborative Research Center TRR110 ``Symmetries and the Emergence of Structure in QCD'' (DFG Project-ID 196253076), the Chinese Academy of Sciences (CAS) under Grant Nos. YSBR-101 and XDB34030000, the EU STRONG-2020 project under the program
H2020-INFRAIA-2018-1 (Grant No. 824093). P.-P.S. also acknowledges the Generalitat valenciana (GVA) for the project with ref. CIDEGENT/2019/015. A.N. is supported by the Slovenian
Research Agency (research core Funding No. P1-0035) and by CAS President’s International Fellowship Initiative (PIFI) (Grant No. 2024PVA0004).

\end{acknowledgments}

\bibliography{X2refs}

\end{document}